# THIRD HARMONIC CAVITY MODAL ANALYSIS

B. Szczesny, I.R.R. Shinton, R.M. Jones,
Cockcroft Institute of Accelerator Science and Technology, Daresbury, UK
School of Physics and Astronomy, University of Manchester, UK

*Abstract*

Third harmonic cavities have been designed and fabricated by FNAL to be used at the FLASH/XFEL facility at DESY to minimise the energy spread along the bunches. Modes in these cavities are analysed and the sensitivity to frequency errors are assessed. A circuit model is employed to model the monopole bands. The monopole circuit model is enhanced to include successive cell coupling, in addition to the usual nearest neighbour coupling. A mode matching code is used to facilitate rapid simulations, incorporating fabrication errors. Curves surfaces are approximated by a series of abrupt transitions and the validity of this approach is examined.

## INTRODUCTION

In operating a FEL an energy variation along the length of an accelerated bunch adversely affects the generation of radiation. The cosine-like fields used to accelerate these bunches at the XFEL and FLASH facility at DESY introduce an energy spread over the length of the bunch. It is desirable to reduce this energy spread by flattening the overall field and this can be achieved by including harmonics of the fundamental frequency of the linac. A single frequency operating at the $n^{th}$ harmonic can be used flatten out the dependence of the energy gain verses phase, by cancelling the second derivative of the fundamental at its peak. In practice the first component in a Fourier expansion is used, namely the 3rd harmonic. This minimises the effect of transverse wakefields, which grow in proportion to the third power in frequency. The overall field profile, with and without the additional harmonic, is illustrated in Fig. 1.

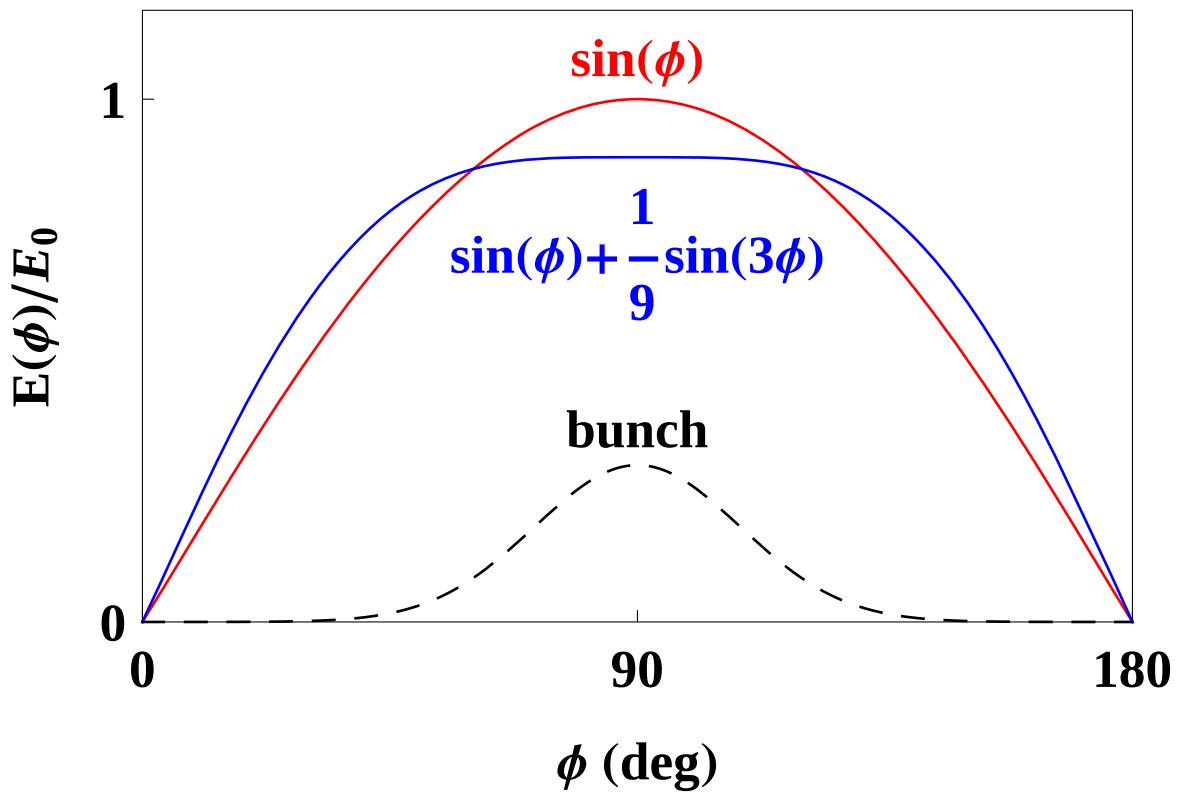

Figure 1: Linearization of rf electric field curvature with third harmonic cavities.

This method will be used at XFEL/FLASH, DESY to compensate for an energy spread across the bunch. A module consisting of four nine-cell cavities operating at 3.9GHz has been designed and built by FNAL (illustrated in Fig. 2) [1].

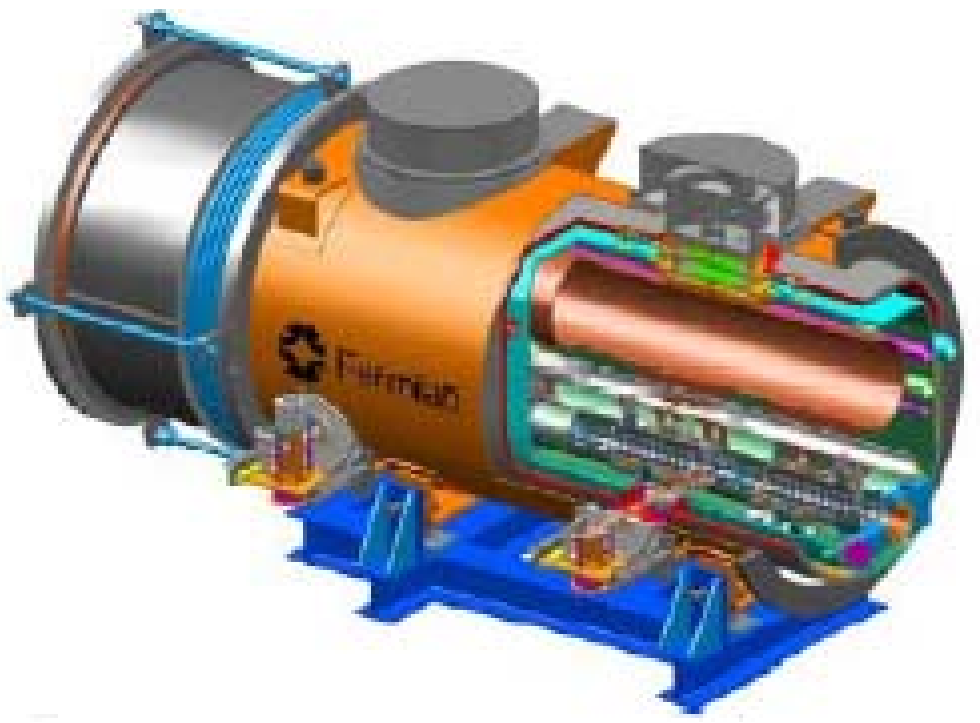

Figure 2: Schematic of a FNAL cryo-module [1] consisting of four 3.9GHz cavities.

This is now in the process of being tested at the DESY Cryo-Module Test Beam (CMTB) facility, prior to installation and use at the FLASH facility. The wakefields in these third harmonic cavities are considerably larger than those in the main accelerating linacs. It is important to damp the modal components of the wakefields and to ensure there are no trapped modes with particularly high R/Q values. An experimental and simulation study is underway to characterise these modes. These modes will also be used to remotely align the beam to the electrical centre of the cavity and to ascertain internal misalignments.

Finally we note that flattening the field also reduces the growth of transverse phase space. The transverse magnetic fields arise from the rate of change of the longitudinal electric field. Thus, flattening the electric field will also result in a reduced magnetic field. Hence the use of a cryo-module of third harmonic cavities will reduce the dilution of both longitudinal and transverse phase space.

The purpose of this paper is to characterise the modes and ascertain their sensitivity to fabrication errors. This paper is organised in three main sections. The first discusses the circuit model applied to monopole modes for several bands. The following section applies a mode matching method, in which curves surfaces are modelled with a series of abrupt transitions in order to rapidly calculate the mode properties in many cavities. The final main section considers the effect of errors on the field profile in these cavities.

## CIRCUIT MODEL OF MONOPOLE BAND

The coupling factor, quality factor and shunt impedance are parameters which depend on the details of the e.m. (electromagnetic) field within individual cavities. In practice these parameters are determined by finite element or finite difference modelling codes or by direct experimental measurement. However, a coupled resonator model (as developed by D. Nagle, E. Knapp and B. Knapp [2]) can used predict these parameters rapidly and accurately. This readily lends itself to including experimental errors and the transient response of the cavity. As is well-known, the properties of the first monopole band can be accurately modelled with a single chain of nearest-neighbour inductively coupled L-C cavities. Here we extend the coupling to include next-nearest neighbour and beyond, in addition to the usual nearest neighbour coupling. The circuit model for the monopole mode is illustrated in Fig 3. Here we have indicated next-nearest neighbour coupling in a set of individual cells, terminated in half-cells to ensure a flat field is obtained in the accelerating π mode. The quality factor of the main cell is given by $Q=2\omega_r L/R$, where $\omega_r/2\pi$ is the cell resonant frequency. In practice the Q is extremely large for superconducting cavities and the resistive loss can effectively be ignored. The equation for these coupled circuits in matrix form can be then written as:

$$\begin{pmatrix} 1 & \kappa_1 & \kappa_2 & 0 & \cdots & 0 \\ \kappa_1/2 & 1 & \kappa_1/2 & \kappa_2/2 & & \vdots \\ \kappa_2/2 & \kappa_1/2 & 1 & \kappa_1/2 & & \vdots \\ 0 & & & \ddots & & \vdots \\ \vdots & & & \kappa_1/2 & 1 & \kappa_1/2 \\ 0 & \cdots & 0 & \kappa_2 & \kappa_1 & 1 \end{pmatrix} \begin{pmatrix} i_0 \\ i_1 \\ \vdots \\ i_n \\ \vdots \\ i_N \end{pmatrix} = \frac{\omega_r^2}{\omega^2} \begin{pmatrix} i_0 \\ i_1 \\ \vdots \\ i_n \\ \vdots \\ i_N \end{pmatrix}$$

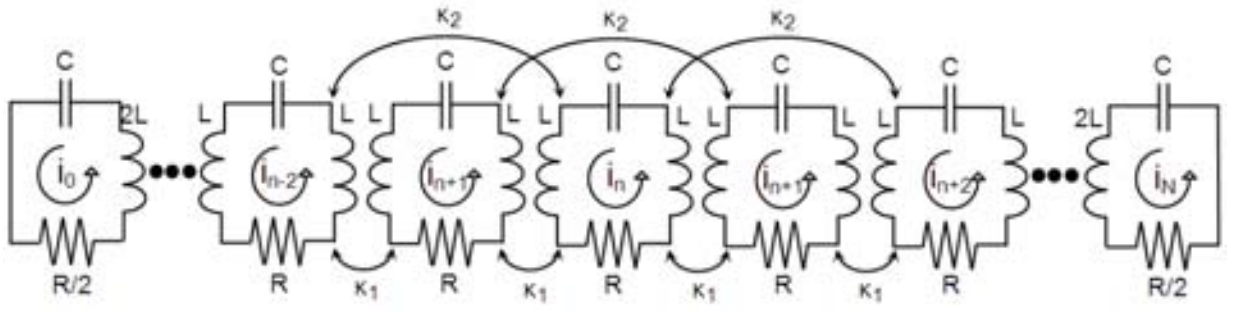

Figure 3: Circuit model of the monopole modes of a series of coupled accelerator cells.

(1),

which can be conveniently written in the condensed form $Hi = \lambda i$. The solution to this matrix equation corresponds to eigenvalues and eigenvectors of the cavity. A single cell from this cavity, subjected to infinite periodic boundary conditions, corresponding to the Floquet condition $i_{n+1} = i_n e^{j\varphi}$ gives the dispersion relation:

$$\left(\frac{\omega_r}{\omega}\right)^2 = 1 + \kappa_1 \cos\varphi + \kappa_2 \cos 2\phi \qquad (2).$$

Adding coupling from successive neighbours modifies this model in a straightforward manner:

$$\left(\frac{\omega_r}{\omega}\right)^2 = 1 + \sum_{n=1}^{n_c} \kappa_n \cos n\varphi \qquad (3)$$

where $n_c$ corresponds to the number of neighbours participating in the coupling. The usual nearest-neighbour coupling corresponds to $n_c=1$. In investigating the influence of errors on the eigenmodes of the cavity each cell is allowed to have a distinct frequency and we use an approximated version of [3]:

$$\begin{pmatrix} 1/\omega_{r0}^2 & \kappa_1/\omega_{r0}^2 & \cdots & 0 \\ \kappa_1/2\omega_{r1}^2 & 1/\omega_{r1}^2 & & \vdots \\ \vdots & & \ddots & \kappa_1/2\omega_{r8}^2 \\ 0 & \cdots & \kappa_1/\omega_{r9}^2 & 1/\omega_{r9}^2 \end{pmatrix} \begin{pmatrix} i_0 \\ \vdots \\ \vdots \\ i_N \end{pmatrix} = \frac{1}{\omega^2} \begin{pmatrix} i_0 \\ \vdots \\ \vdots \\ i_N \end{pmatrix} \qquad (4)$$

We utilised the circuit model described by Eq. (3) to fit the dispersion curves of higher monopole bands. The first eight monopole band dispersion curves were simulated with HFSS v11 for a cavity consisting of nine cells with beam pipes attached and terminated in either electric-electric (E-E) or magnetic-magnetic (H-H). Provided the mode is well-contained within the cavity the eigenmodes will be insensitive to the terminating boundary conditions. However, modes which are able to propagate outside the cavity, or modes with significant evanescent tails will be sensitive to the termination conditions. In order the avoid sensitivity to the latter condition we ensured the terminations were placed several wavelengths beyond the end-cells.

Several of the results of these simulations were compared with Superfish [3] simulations. However, even after choosing a variety of excitation points for the drive current, we were not able to obtain all modes with Superfish simulations as the modes were particularly close for a number of bands.

The first monopole band is below the cut-off of the beam pipes and hence these modes are confined within the cavity. Other bands have the potential for modes to propagate out of the cavity. In order to accurately simulate the monopole modes of the cavity a 10° slice was used with H-H symmetry planes. The results of this simulation are displayed in Fig. 4, together with those predicted by the circuit model. The parameters used in fitting the circuit model are identified by the black points in Fig. 4. The standard dispersion relation ($n_c=1$) requires two points from each fitted curve and is displayed in Fig. 4(a). In each band we utilise two points: one closest to 0 mode and another close to the π mode. It is clear that, as expected, the first band is well-predicted by the circuit model. However, there is a significant discrepancy between that predicted by the circuit model and that obtained from the HFSS v11 [4] simulations for higher bands. Adding coupling from two additional neighbours ($n_c=3$), using two additional points to obtain the two additional parameters, improves the prediction significantly and this is displayed in Fig 4(b). This implies the fields couple beyond their nearest neighbours.

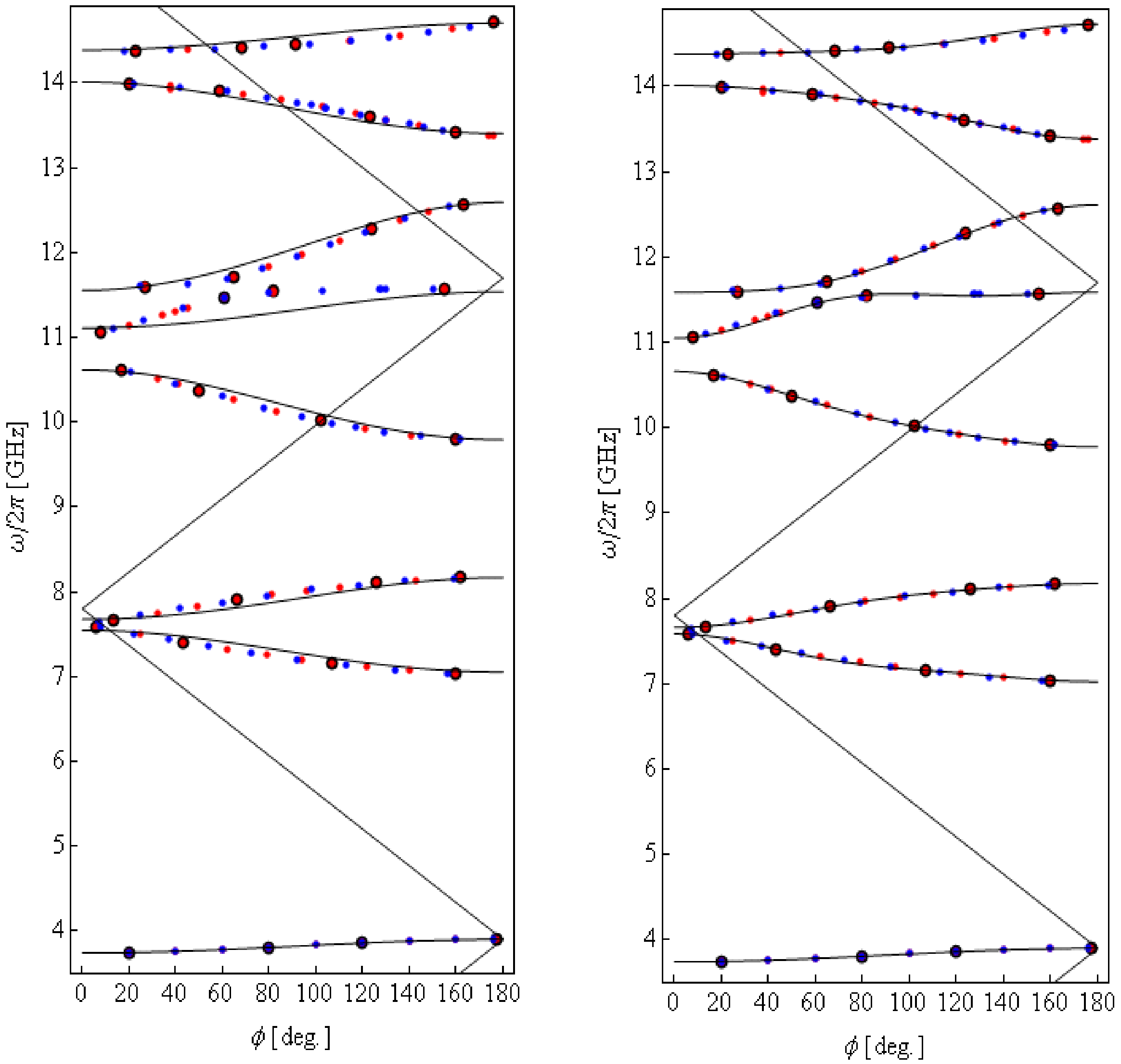

Figure 4: Dispersion curves of complete nine-cell cavity with attached beam tubes. HFSS v11 is used in the eigenmode module with E-E terminations (blue points) and H-H terminations (red points). Curves from the standard circuit model, $n_c$ =1, are also shown in (a) and the extended coupling model, $n_c$ = 3, in (b). The points used in determining the circuit model parameters are indicated by the black points.

Many of these modes in the higher bands propagate out of the cavity and in this case simulations with radiation boundary conditions are more appropriate. Taking this into account properly with suitable finite element or finite difference simulations remains the subject for future work. Two typical representative modes from the second band, both a mode contained within the cavity and a multi-cavity mode, are illustrated in Fig. 5.

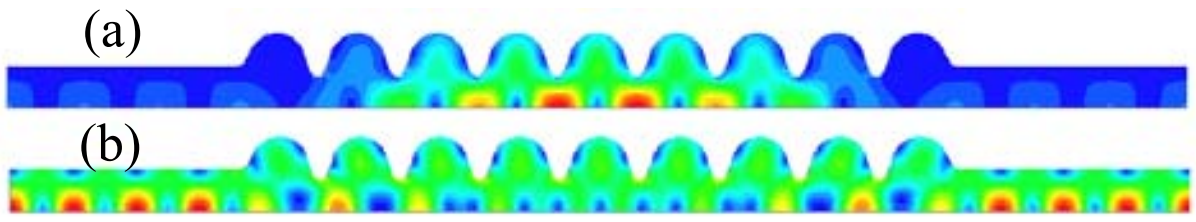

Figure 5: HFSS v11 simulations indicating (a) a trapped mode at 7.04GHz (the first mode of the second monopole band) and (b) a multi-cavity mode at 7.58GHz (the ninth mode of the second monopole band).

## MODE MATCHING METHOD APPLIED TO MULTICELL SC CAVITITES

In order to calculate the sensitivity of the cavity R/Qs and mode frequencies to fabrication errors we utilised the scattering matrix code Smart2d [5]. These parameters are rapidly determined from simulations of the impedance of the overall cavity. This code propagates modes through sharp wide (W) to narrow (N) transitions in waveguides.

Clearly the third harmonic SC cavities are by no means abrupt discontinuities. Thus, we slice up each curved surface, the iris for example, into a series of WN and NW transitions. Prior to simulating a full nine-cell cavity we first verified the accuracy of computing the scattering matrix for a single cell. We chose the middle cell of the cavity and compared an accurate HFSS simulation. In this case HFSS is operated in the driven mode module. It is convenient to express the S-matrix in terms of three parameters, the angles: θ, ϕ and dϕ [6]. The S-matrix is represented in the form:

$$\begin{pmatrix} S_{11} & S_{12} \\ S_{21} & S_{22} \end{pmatrix} = -\begin{pmatrix} \cos\theta e^{jd\phi} & j\sin\theta \\ j\sin\theta & \cos\theta e^{-jd\phi} \end{pmatrix} e^{j\phi} \quad (5)$$

In this representation the S-matrix is guaranteed to be unitary. The objective of these simulations is to ascertain the minimum number of slices which are sufficient to accurately model the e.m. fields across the curved surfaces. A representative set of simulations at the π mode frequency of 3.9 GHz is shown in Table 1. The parameter dϕ =0 is a reflection of the symmetry of the transition being simulated ($S_{11} = S_{22}$). Twenty segments accurately simulates the S-matrix of the surface to within a few percent. Armed with this knowledge on the number of sliced required, we used the impedance module of the Smart2d code to calculate the impedance of the complete 9-cell cavity. The results of simulations of the real part of the impedance, with 20 slices for each curved transition, are displayed in Fig. 6. Here we added a small imaginary frequency, corresponding to a loss component in order to broaden the characteristic resonances. Several resonances overlap in the vicinity of the regions in which the bands are particularly close (close to 0 phase advance for the second and third bands, for example). In these regions it is difficult to pameterise the resonances. The first band is well-characterised however. A non-linear least square error fit, with a series of Lorentzian functions [7]

$$L_n(\omega) = \frac{2Q_n}{\pi\omega_n} \frac{K_n}{1 + 4Q_n^2 (\omega/\omega_n - 1)^2} \quad (6)$$

is made to the data resulting from Smart2D simulations. The n[th] resonance is fitted by varying three parameters: $Q_n$, $K_n$, and $\omega_n$. Provided the modes are well-separated the Q corresponding to the single cell result is also obtained for the modal value. Closely spaced modes share Qs and consequently the cell values are modified. The $R_n/Q_n$ ($=K_n/\omega_n$) and mode frequency $\omega_n/2\pi$ achieved from these fits for the nth mode are compared with simulations made with MAFIA2D [8] and HFSS v11 [9] in Fig. 7. Bearing in mind the way in which the curvature of the surfaces has been grossly approximated with sharp transitions the agreement between modes frequencies and R/Qs is encouraging.

Table 1: Parameterisation of S-matrix calculated with the Smart2d code as a function of the number of conjoined sharp transitions (Number of slices per half length).

| Slice | 10 | 20 | 40 | HFSS |
|---|---|---|---|---|
| $\theta$ | 0.2799 | 0.2897 | 0.2983 | 0.2852 |
| $\phi$ | 2.7630 | 2.8022 | 2.8356 | 2.8540 |
| $d\phi$ | 0 | 0 | 0 | 0.0001 |

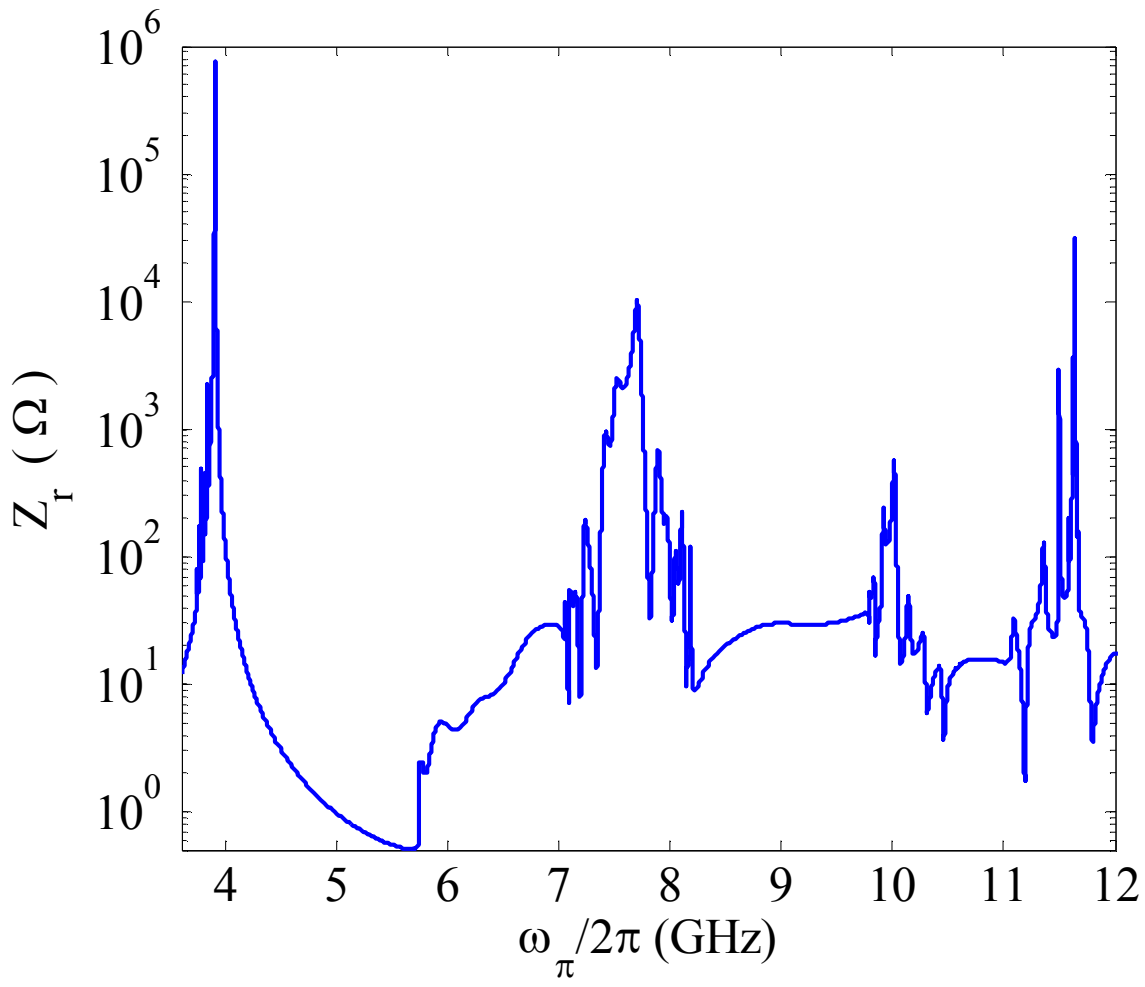

Figure 6: Real part of the beam impedance $Z_r$ as a function of frequency for a full 9cell cavity with beam pipes attached. Bands 1 to 5 are shown.

In order to ascertain the sensitivity of these cavities to manufacturing errors we made similar simulations in which we added uniformly distributed random errors. with an rms of 100μm to several geometrical parameters. Errors of this magnitude may result from the many stages entailed in the fabrication of these cells. However the relevance of these large errors introduced needs to be confirmed and it may be an overestimate of the overall fabrication errors. The results of this simulation with twenty cavities subjected errors in seven randomly varied geometric parameters (iris radius, equator radius, half cell length, equator horizontal axis, equator vertical axis, iris horizontal axis and iris vertical axis) is illustrated in Fig. 8. The original, unperturbed structure R/Qs are shown with blue points and the rms of the R/Qs of all structures is indicated by the red points. Here it is notable that those modes close to the accelerating π mode suffer significant frequency and R/Q shifts. Indeed the main accelerating mode R/Q is degraded from more than 375Ω to ~260 Ω.

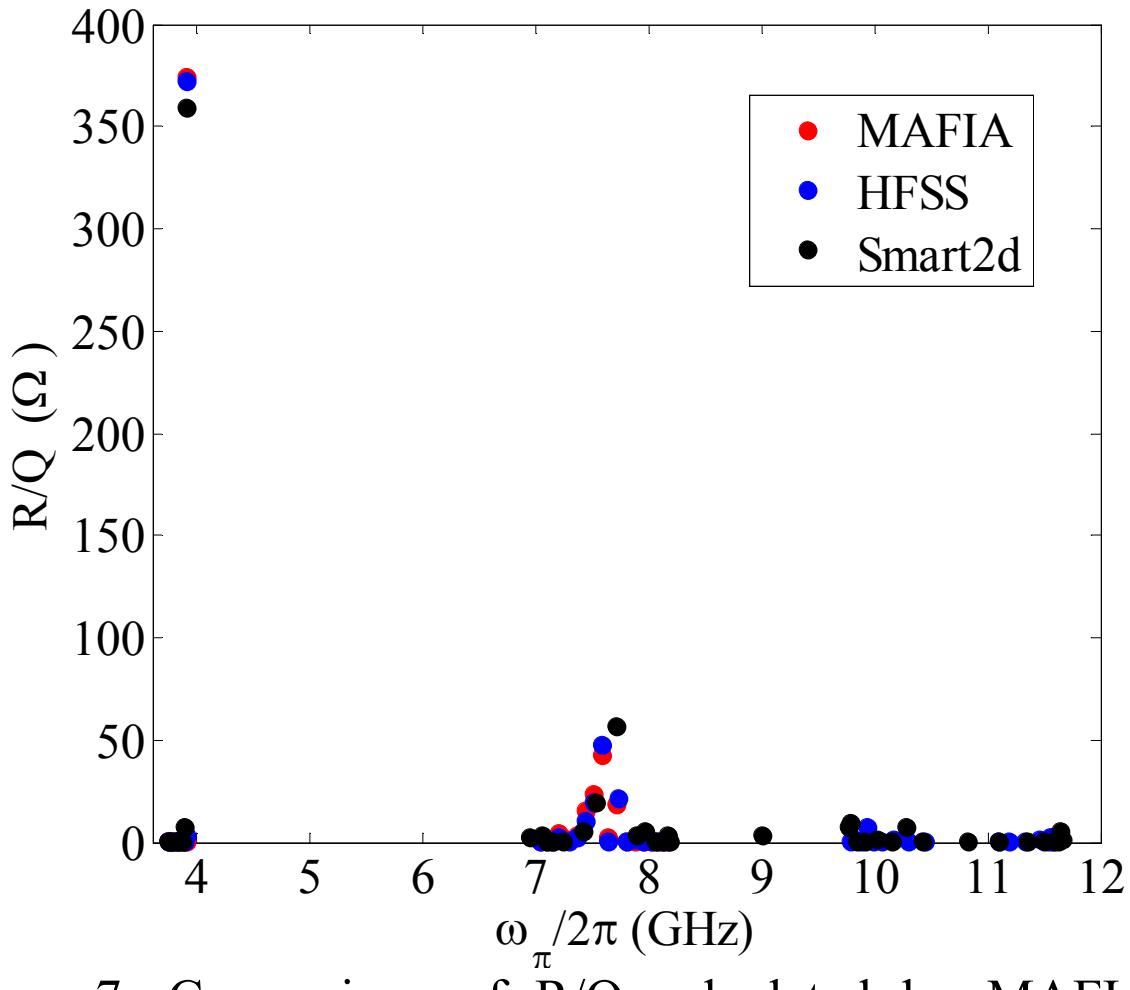

Figure 7: Comparison of R/Q calculated by MAFIA, HFSS v11 and SMART2D for bands 1 to 5 of a full 9cell cavity with beam pipes attached.

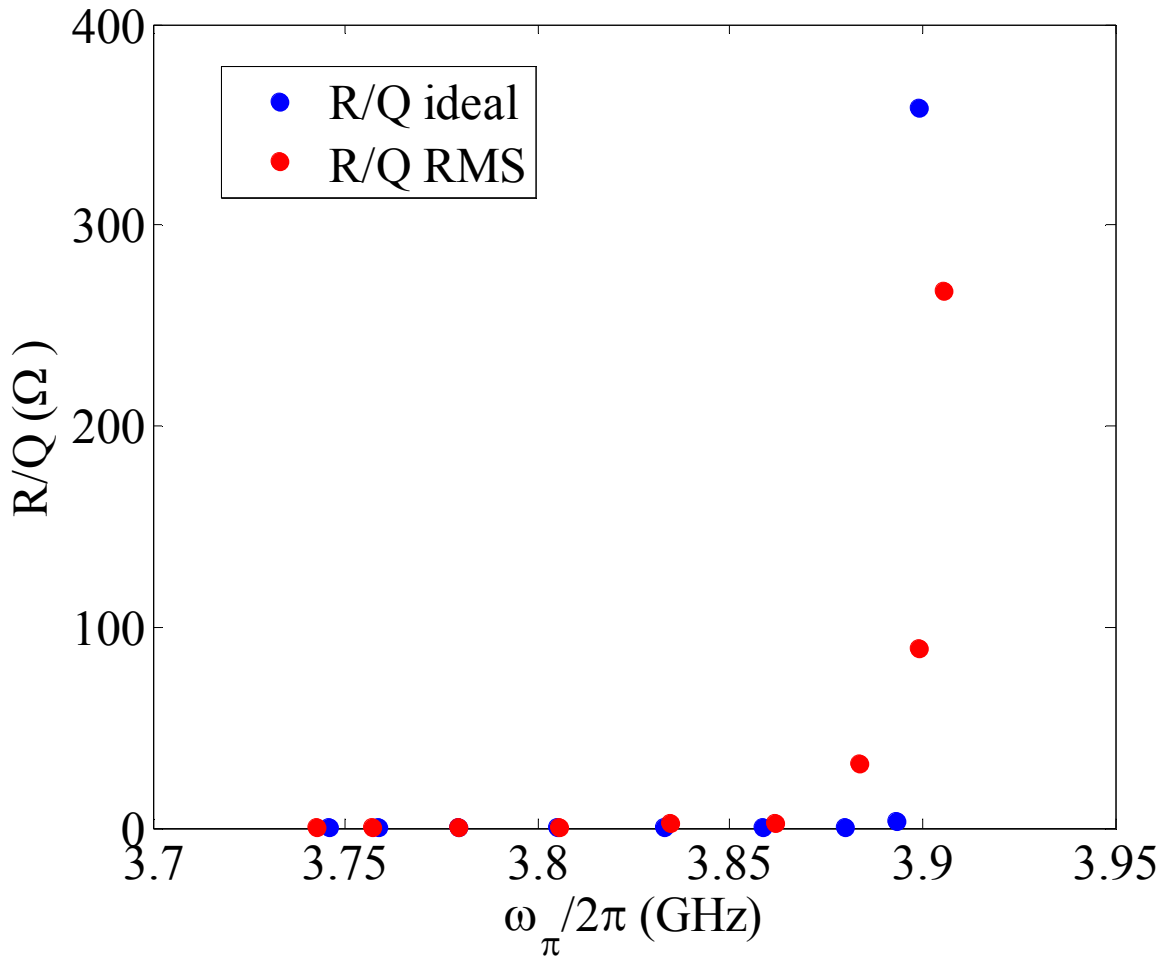

Figure 8: Comparison of the R/Qs for the first monopole band for an idealised perfect structure and the rms of 20 different simulations. Here all seven geometric cell parameters for each cell were randomly varied within an rms tolerance of ±100μm.

## INFLUENCE OF RANDOM ERRORS ON FIELD FLATNESS

Random errors are introduced into each of the 9-cells and the field profile is calculated in a manner similar to that performed in [10]. In order to tune the field back to the perfectly flat original state we allow two tuners in either end of the cavity to be used. We seek to minimise the "cost" function, determined as the sum of the squares of the difference of the field from the perfectly flat original state. A computer code was written in Mathematica [8] for this purpose. We have investigated the effect of random errors with a uniform distribution that have an rms of 5MHz. The resulting field is flat to within 8% for a 5MHz error. In addition one of these simulations that result from this automated tuning process is also shown for an rms error of 5MHz in Fig. 9. The field error, prior to tuning has an rms deviation from unity of 29.9% ± 16.4% and after tuning it is 8.3% ± 3.3%. The necessary tuning required is 19MHz and -10.6MHz in the first and last end cells respectively.

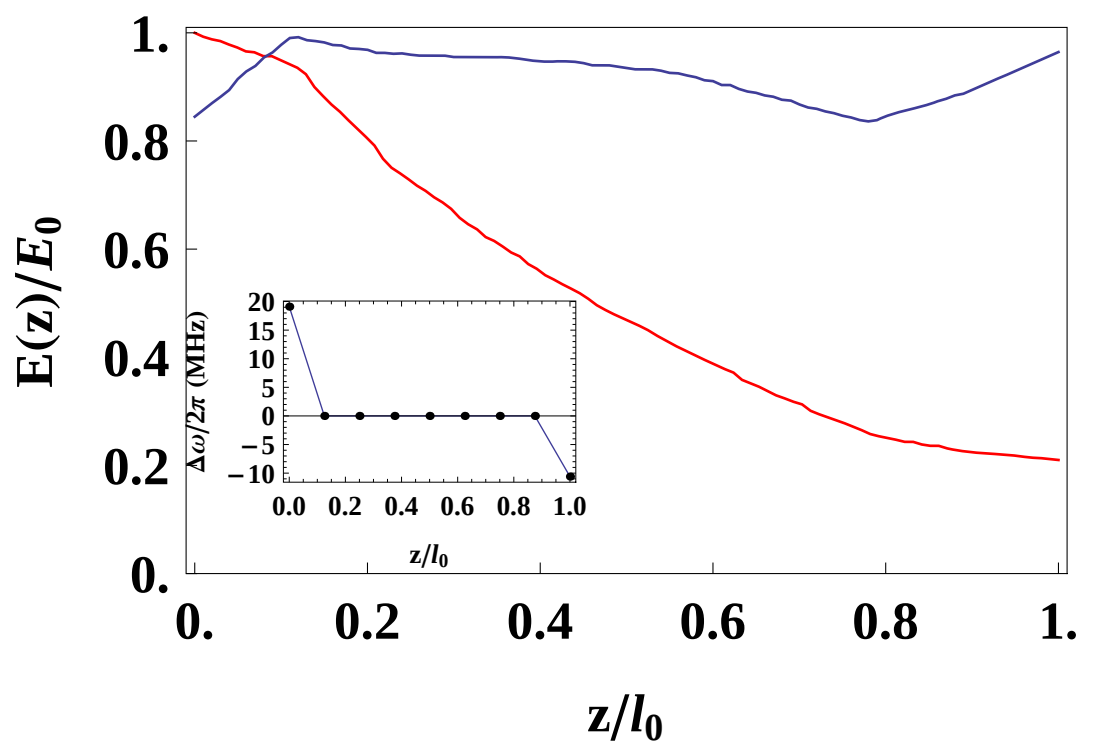

Figure 9: The Field along the cavity of length $l_0$, prior to tuning, is indicated by the red curve. A correction algorithm, entailing tuning the end cells only results in a significant improvement in the field flatness and this is indicated by the blue curve. The corresponding correction frequencies Δω/2π are inset.

## FINAL REMARKS

The addition of successive neighbour coupling results in a significant improvement in the agreement between the predicted and full finite element modelling of the dispersion curves of the monopole band of the third harmonic FNAL cavities fabricated for FLASH. This should prove a useful design tool in analysing the eigenmodes of the complete structure. Initial simulations have been conducted on the sensitivity of the cavity to fabrication errors.

## ACKNOWLEDGEMENTS

We have benefited from discussions at the weekly Manchester Electrodynamics and Wakefields (MEW) meeting held at Cockcroft Institute, where these results were first presented. This research has received funding from the European Commission under the FP7 Research Infrastructures grant agreement no.227579. This document contains material, which is the copyright of certain EuCARD beneficiaries and the European Commission, and may not be reproduced or copied without permission. The information herein does only reflect the views of its authors and not those of the European Commission. The European Commission and the EuCARD beneficiaries do not warrant that the information contained herein is capable of use, or that use of the information is free from risk, and they are not responsible for any use that might be made of data appearing herein